\documentclass[12pt]{article}
\pdfoutput=1
\usepackage{draft,tikz-cd}

\usepackage{tikz,tikz-3dplot}
\usepackage{pgfplots}
\pgfplotsset{compat=1.17}
\usetikzlibrary{shapes,arrows,cd,chains,decorations.markings,decorations.pathmorphing,calc}
\tikzset{
	->-/.style args={#1rotate#2}{decoration={markings, mark=at position #1 with {\arrow[scale=1.5,rotate = #2 ]{stealth}}}, postaction={decorate}}
}
\tikzstyle{GraphNode}=[circle, draw=black, fill=black, inner sep=2pt, minimum size=5pt]
\tikzstyle{GraphEdge}=[black]
\pgfmathsetmacro{\gS}{1}

\begin{document}

\begin{titlepage}

\begin{center}

\title{Violation of S-duality in classical $Q$-cohomology}

\author{Chi-Ming Chang$^{a,b,c}$ and Ying-Hsuan Lin}

\address{${}^a$Yau Mathematical Sciences Center (YMSC), Tsinghua University, Beijing, China}

\address{${}^b$Beijing Institute of Mathematical Sciences and Applications (BIMSA) Beijing, China}

\address{${}^c$Peng Huanwu Center for Fundamental Theory, Hefei, Anhui, China}

\email{cmchang@tsinghua.edu.cn,  yhlin@alum.mit.edu}

\end{center}

\vfill

\begin{abstract}

We study the cohomology of a chiral supercharge $Q$ in the $\mathcal{N}=4$ super-Yang-Mills (SYM) theory at tree level. The cohomology classes correspond one-to-one to the $\frac1{16}$ Bogomol'nyi-Prasad-Sommerfield (BPS) states at one-loop. We argue that monotone classes on the Coulomb branch respect the S-duality between the theories with $\mathrm{SO}(2N+1)$ and $\mathrm{USp}(2N)$ gauge groups, but find an explicit example of a pair of cohomology classes that ``violate'' the S-duality in the sense that the tree-level $Q$-cohomologies are not isomorphic between the neighborhoods near the two free points. Within this pair, one is a fortuitous class and the other is a monotone chiral ring element. Assuming the non-perturbative validity of S-duality, our results disprove a long-standing conjecture on the one-loop exactness of the $\frac1{16}$-BPS spectrum (including the $\frac1{8}$-BPS chiral ring spectrum) in the $\mathcal{N}=4$ SYM. Mathematically, this shows that, the relative Lie algebra cohomology $H^\bullet(\mathfrak{g}[A],\mathfrak{g})$ is generally not graded-isomorphic to $H^\bullet({}^L\mathfrak{g}[A],{}^L\mathfrak{g})$, where $\mathfrak{g}$ and ${}^L\mathfrak{g}$ are a pair of Langlands dual Lie algebras and $A=\mathbb{C}[z^+,z^-]\otimes\Lambda(\theta_1,\theta_2,\theta_3)$.

\end{abstract}

\vfill

\end{titlepage}

\section{Introduction}

The study of supercharge $Q$-cohomology in supersymmetric field theories dates back to Witten's work \cite{Witten:1982df}, where the Euler characteristic of the $Q$-cohomology, namely the Witten index, was argued to be independent of couplings. However, how the $Q$-cohomology itself depends on the couplings remains an open question. It was conjectured implicitly in \cite{Kinney:2005ej,Berkooz:2006wc} and explicitly in \cite{Grant:2008sk} that the spectrum of $Q$-cohomology classes in the ${\cal N}=4$ super-Yang-Mills (SYM) is classically (tree-level) exact. This non-renormalization conjecture opened a window for studying the microstates of black holes in the gravity dual of the ${\cal N}=4$ SYM at strong 't Hooft coupling, via constructing and manipulating the $Q$-cohomology classes at weak coupling, and motivated a series of more recent works \cite{Chang:2013fba,Chang:2022mjp,Choi:2022caq,Choi:2023znd,Budzik:2023xbr,Chang:2023zqk,Budzik:2023vtr,Chang:2023ywj,Choi:2023vdm,Gaiotto:2024gii,Chang:2024zqi,Chen:2024oqv,deMelloKoch:2024sdf,Chang:2024lxt,deMelloKoch:2024pcs,Chang:2025rqy,deMelloKoch:2025ngs,deMelloKoch:2025cec,Hughes:2025car,Gadde:2025yoa}.

The ${\cal N}=4$ SYM theory enjoys S-duality, which maps the theory with gauge group $G$ and complexified gauge coupling 
\ie
\tau=\frac{\theta}{2\pi}+\frac{4\pi i}{g_{\rm YM}^2}
\fe 
to the theory with gauge group ${}^L G$ (the Langlands dual of $G$) and complexified gauge coupling $-1/\tau$. Since the dual pair of theories admit weak coupling descriptions near two different points on the space of couplings, S-duality provides a powerful tool for testing the non-renormalization conjecture. The pair with gauge groups $\mathrm{SU}(N)$ and $\mathrm{PSU}(N)$ does not allow for any non-trivial check because the $Q$-cohomology depends only on the Lie algebra of the gauge group. To perform nontrivial checks, this paper considers the gauge groups $\mathrm{SO}(2N+1)$ and $\mathrm{USp}(2N)$. A fortuitous class in the $\mathrm{SO}(7)$ theory was recently constructed in \cite{Gadde:2025yoa}, where some hypotheses concerning S-duality were presented in \cite[Section~4.4]{Gadde:2025yoa} even though they did not explicitly study the $\mathrm{USp}(6)$ theory.\footnote{The techniques of \cite{Gadde:2025yoa} are specialized to the Berenstein-Maldacena-Nastase (BMN) sector \cite{Berenstein:2002jq}, but S-duality does not logically have to map between the BMN sectors of the dual theories.}

In Section~\ref{sec:formalism}, we review the formalism for computing the tree-level $Q$-cohomology and argue for a one-to-one correspondence between the cohomology classes on the Coulomb branch of the $\mathrm{SO}(2N+1)$ and $\mathrm{USp}(2N)$ theories; these Coulomb branch cohomology classes form a subset of the monotone classes. In Section~\ref{sec:results}, we present our main result, which is a disproof of the non-renormalization conjecture by two additional cohomology classes in the $\mathrm{SO}(7)$ theory relative to the $\mathrm{USp}(6)$ theory: one fortuitous and one monotone.

\section{Formalism}
\label{sec:formalism}

Following \cite{Chang:2013fba}, consider an adjoint-valued fermionic superfield $\Psi$ vanishing at the origin of the superspace $\mathbb{C}^{2|3}$:
\ie
\left.\Psi(z^+,z^-,\theta_1,\theta_2,\theta_3)\right|_{z^\pm=\theta_{1,2,3}=0}=0.
\fe
The supercharge $Q$ acts by
\ie\label{Qaction}
Q(\Psi)=\Psi^2,
\fe
and extends to composites via the graded Leibniz rule:
\ie
Q(\cO_1\cO_2)=Q(\cO_1)\cO_2+(-1)^{|\cO_1|}\cO_1Q(\cO_2),
\fe
where $|\cO|$ denotes the fermion parity of $\cO$. The cochain complex computing the tree-level $Q$-cohomology is spanned by the multi-trace operators built from derivatives of $\Psi$, inserted at the origin $z^\pm=\theta_i=0$, modulo trace relations, and graded by the number of $\Psi$ insertions. As was shown in \cite{Chang:2013fba}, this tree-level $Q$-cohomology is canonically isomorphic to the relative Lie algebra cohomology
\ie
H^n(\mathfrak g[A],\mathfrak g),
\fe
via identifying the above operator complex with the relative Chevalley–Eilenberg complex $C^n(\mathfrak g[A],\mathfrak g)$, where $\mathfrak g$ is the Lie algebra of the gauge group, $A=\mathbb{C}[z^+,z^-]\otimes\Lambda(\theta_1,\theta_2,\theta_3)$, and the cohomological degree $n$ is the number of $\Psi$ insertions. In addition to $n$, a cohomology class carries five more gradings $(n_+,n_-,n_1,n_2,n_3)$: $n_\pm$ counts the number of insertions of $\partial_{z^\pm}$, and $n_i$ (for $i=1,2,3$) counts $\partial_{\theta_i}$. Note that only five linear combinations $(J_1, J_2, q_1, q_2, q_3)$ are conserved charges, and only four appear in the graded Witten index.

Let us consider the one-form
\ie
d\Psi=dz^{\A}\,\partial_{z^{\A}}\Psi+d\theta_i\,\partial_{\theta_i}\Psi.
\fe
The supercharge acts on $d\Psi$ by
\ie\label{eqn:Q_dPsi}
Qd\Psi=[\Psi,d\Psi],
\fe
and it follows that the single-trace operators
\ie
\Tr\left[(d\Psi)^n\right]
\fe
are $Q$-closed. Expanding at the origin, we obtain
\ie
\sum_{\vec v} {\cal O}_{\vec v} \prod_{\alpha=\pm} (z^\alpha)^{p_\alpha} \prod_{i=1,2,3} (\theta_i)^{p_i}\prod_{\beta=\pm} (dz^\beta)^{m_\beta} \prod_{j=1,2,3} (d\theta_j)^{m_j},
\fe
where $\vec v=(p_{\A},p_i,m_{\B},m_j)$ and
\ie\label{eqn:single_trace_coho}
{\cal O}_{\vec v}=\prod_{\A=\pm}\prod_{i=1}^3\partial_{z^{\A}}^{p_{\A}}\partial^{p_i}_{\theta_i}\Tr\bigg[\underbrace{\prod_{\beta=\pm}(\partial_{z^{\B}}\Psi)^{m_{\B}}\prod_{j=1}^3(\partial_{\theta_j}\Psi)^{m_i}}_{\text{(anti-)symmetrized}}\bigg]\bigg|_{z^\pm=\theta_{1,2,3}=0}.
\fe
multi-trace $Q$-closed operators arise as products
\ie\label{eqn:multi-grav_form}
{\cal O}_{\vec v_1}\cdots {\cal O}_{\vec v_k},
\fe
which are redundant due to trace relations. After imposing the trace relations, the multi-trace operators \eqref{eqn:multi-grav_form} serve as representatives of the monotone $Q$-cohomology classes \cite{Chang:2013fba,Chang:2024zqi}. Beyond the monotone sector, there exist ``fortuitous'' classes that cannot be represented by \eqref{eqn:multi-grav_form}; the first example was found in the theory with gauge group $\mathrm{SU}(2)$ \cite{Chang:2022mjp}.

In the following, we focus on the gauge groups $\mathrm{SO}(2N+1)$ and $\mathrm{USp}(2N)$. In the $\mathrm{SO}(2N+1)$ theory, the adjoint superfield $\Psi$ is antisymmetric,
\ie
\Psi^T=-\Psi,
\fe
while in the $\mathrm{USp}(2N)$ theory, $\Psi$ satisfies the symplectic condition
\ie
\Omega\Psi+\Psi^T\Omega=0, \quad {\rm with}\quad
\Omega=\begin{pmatrix}
0 & I_N\\
-I_N & 0
\end{pmatrix},
\fe
where $I_N$ is the $N\times N$ identity matrix. Gauge transformations act by conjugation,
\ie
\Psi\to M^{-1}\Psi M,
\fe
with $M$ independent of $z^\pm$ and $\theta_i$. For $\mathrm{SO}(2N+1)$, $M$ is real and orthogonal, while for $\mathrm{USp}(2N)$, $M$ is unitary and symplectic, i.e.,
\ie
M^T\Omega M=\Omega.
\fe

\subsubsection*{An isomorphism between the cohomologies on the Coulomb branch}

Consider the theory at a generic point on the Coulomb branch, where the gauge group is broken to its maximal torus $\mathrm{U}(1)^N$.
In a convenient basis, for the $\mathrm{SO}(2N+1)$ theory the Cartan-valued superfield $\Psi$ takes the block off-diagonal form
\ie\label{eqn:SO_commuting}
\Psi=(\sigma_2\otimes\Psi_D)\oplus \begin{pmatrix}0\end{pmatrix}=\begin{pmatrix}
0 & -i\Psi_D & 0 \\
i\Psi_D & 0 & 0 \\
0 & 0 & 0 
\end{pmatrix},
\fe
while for the $\mathrm{USp}(2N)$ theory, we have
\ie\label{eqn:Sp_commuting}
\Psi=\sigma_3\otimes\Psi_D=\begin{pmatrix}
\Psi_D & 0 \\
0 & -\Psi_D
\end{pmatrix}.
\fe
Here $\Psi_D$ is an $N$-dimensional diagonal matrix, $\sigma_{2,3}$ are Pauli matrices, and $I_N$ denotes the $N$-dimensional identity matrix. The Coulomb branch cohomology is defined by ``(super)abelianization'': substituting the matrix \eqref{eqn:SO_commuting} or \eqref{eqn:Sp_commuting} into the cohomology representatives at the origin of the Coulomb branch.
This procedure is consistent at the cohomological level because the $Q$-exact operators must vanish as they are proportional to $\Psi^2 = \frac12 \{\Psi, \Psi\}$ due to the $Q$-action \eqref{Qaction} and the Leibniz rule.\footnote{Note that the Coulomb branch cohomology is different from the cohomology of the free $\mathrm{U}(1)^N \rtimes W$ gauge theory at infinite distance on the Coulomb branch ($W$ is the Weyl group), as many non-BPS states become BPS in this limit.
}

Since the last row and column in \eqref{eqn:SO_commuting} are zero, we may restrict to the nontrivial $2N\times 2N$ block. In this block, the two matrices in \eqref{eqn:SO_commuting} and \eqref{eqn:Sp_commuting} are related by conjugation:
\ie\label{eqn:M_matrix}
L^{-1}\begin{pmatrix}
0 & -i\Psi_D \\
i\Psi_D & 0 
\end{pmatrix} L = \begin{pmatrix}
\Psi_D & 0 \\
0 & -\Psi_D
\end{pmatrix} ,\quad L=\begin{pmatrix}
-iI_N & iI_N \\
I_N & I_N
\end{pmatrix},
\fe
which is neither an $\mathrm{SO}(2N+1)$ nor a $\mathrm{USp}(2N)$ gauge transformation, but it preserves the multi-trace structure of operators. 
We conclude that the Coulomb branch cohomologies in the two ${\cal N}=4$ SYM theories are isomorphic.\footnote{The same argument was independently formulated in \cite[Section~4]{Gadde:2025yoa}.}

Finally, the Coulomb branch cohomology embeds into the monotone cohomology. Indeed, the fortuitous cohomology is identified with a subspace of the cohomology of trace relations \cite{Chang:2024zqi}, and all trace relations vanish identically on the Coulomb branch. It was implicitly conjectured in \cite{Choi:2023znd} that the Coulomb branch cohomology exhausts the monotone cohomology, a consequence of which is that the monotone cohomologies of the $\mathrm{SO}(2N+1)$ and $\mathrm{USp}(2N)$ theories would have to be isomorphic.\footnote{We thank Jaehyeok Choi and Seok Kim for helpful discussions on this point.}
A counterexample to this conjecture will be presented momentarily.

\section{Results and discussion}
\label{sec:results}

To further test the non-renormalization conjecture, we include both the multi-trace monotone sector and the fortuitous sector. As no closed-form formula is known for the latter, we proceed by explicit construction. 

\subsubsection*{Summary of the results}
Focusing on the simplest S-dual pair: $\mathrm{SO}(7)$/$\mathrm{USp}(6)$, we constructed the cohomology classes up to level
\ie
L=3n_{+}+3n_{-}+2n_{1}+2n_{2}+2n_{3}=18,
\fe
and found that the cohomology counts agree in all charge sectors except two, as summarized in Table~\ref{tab:results}.\footnote{The result in the first column of Table~\ref{tab:results} was independently obtained by \cite[Section~4]{Gadde:2025yoa}.}
\begin{table}[H]
\begin{center}
\begin{tabular}{|c|c|c|c|c|}
\hline
$(J_1,J_2,q_1,q_2,q_3)$ & \multicolumn{2}{c|}{$(\frac12,\frac12,\frac52,\frac 52,\frac52)$} & \multicolumn{2}{c|}{(0,0,3,3,3)}
\\\hline
$(n_+,n_-,n_1,n_2,n_3,n)$ & \multicolumn{2}{c|}{(0,0,3,3,3,8)} & \multicolumn{2}{c|}{(1,1,2,2,2,8)}
\\\hline
gauge group & $\mathrm{SO}(7)$ & $\mathrm{USp}(6)$ & $\mathrm{SO}(7)$ & $\mathrm{USp}(6)$
\\\hline
all states & 903 & 903 & 826 & 825
\\\hline
non-$Q$-closed states & 220 & 221 & 0 & 0
\\\hline
$Q$-exact states & 559 & 559 & 741 & 741
\\\hline
monotone classes & 123 & 123 & 85 & 84
\\\hline
fortuitous classes & 1 & 0 & 0 & 0
\\\hline
\end{tabular}
\end{center}
\caption{\label{tab:results} Comparison of cohomology counts between $\mathrm{SO}(7)$ and $\mathrm{USp}(6)$ at level $L=18$ in the two exceptional charge sectors.}
\end{table}
At level $18$, the $\mathrm{SO}(7)$ theory contains two additional cohomology classes relative to $\mathrm{USp}(6)$: a fermionic fortuitous class and a bosonic monotone class. 
These two classes contribute with opposite statistics and cancel in the superconformal index, as is consistent with the equality of the index in the $\mathrm{SO}(2N+1)$ and $\mathrm{USp}(2N)$ theories. 

\subsubsection*{Discussions}

Our result has several immediate implications. First, it rules out an isomorphism between the classical (tree-level) \(Q\)-cohomologies of the \(\mathrm{SO}(2N{+}1)\) and \(\mathrm{USp}(2N)\) theories. Mathematically, this means that the relative Lie algebra cohomologies \(H^\bullet(\mathfrak{g}[A],\mathfrak{g})\) and \(H^\bullet({}^L {\mathfrak g}[A],{}^L {\mathfrak g})\) are generally not graded-isomorphic, where \(\mathfrak{g}\) and \({}^L {\mathfrak g}\) form a Langlands dual pair of Lie algebras. This leaves two possibilities: either S-duality is not exact in \({\cal N}=4\) SYM, or the non-renormalization conjecture that the \(Q\)-cohomology is exact at tree level is false. 

The former is logically possible but highly unlikely; the latter would imply that the additional BPS pair in \(\mathrm{SO}(7)\) acquires an anomalous dimension from higher-loop or non-perturbative effects and is lifted (ceases to be BPS). Loop corrections to the \(Q\)-cohomology in the \({\cal N}=1\) SYM were studied by \cite{Budzik:2023xbr} using the holomorphic twist formalism, which applies equally well to the \({\cal N}=4\) SYM. Moreover, it was argued that for cohomology classes with finite charges, the loop expansion truncates at a finite order. Consequently, a finite computation suffices to determine whether the additional BPS pair in \(\mathrm{SO}(7)\) is lifted by perturbative or non-perturbative effects.

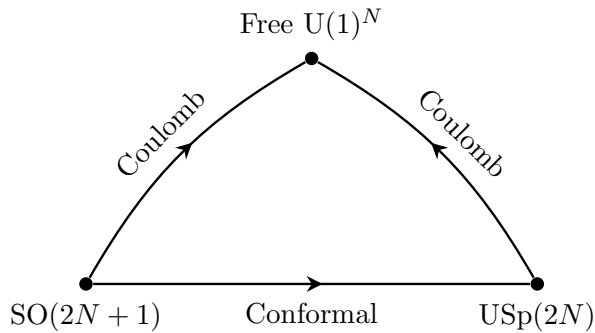
\begin{figure}
\centering
\begin{tikzpicture}[every node/.style={font=\small}, point/.style={circle,fill,inner sep=1.8pt}]
	\node[point,label=below:{$\mathrm{SO}(2N+1)$}] (A) at (0,0) {};
	\node[point,label=below:{$\mathrm{USp}(2N)$}] (B) at (6,0) {};
	\node[point,label=above:{Free $\mathrm{U}(1)^N$}] (C) at (3,3) {};

	\draw[
		line width=0.9pt,
		postaction={decorate},
		decoration={markings, mark=at position 0.525 with {\arrow{Stealth[length=6pt,width=6pt]}}}
	]
	(A) -- node[midway, below, yshift=-3pt, sloped]{Conformal} (B);

	\draw[
		line width=0.9pt,
		postaction={decorate},
		decoration={markings, mark=at position 0.55 with {\arrow{Stealth[length=6pt,width=6pt]}}}
	]
	(A) to[out=60,in=210] node[midway, above, yshift=3pt, sloped] {Coulomb} (C);

	\draw[
		line width=0.9pt,
		postaction={decorate},
		decoration={markings, mark=at position 0.55 with {\arrow{Stealth[length=6pt,width=6pt]}}}
	]
	(B) to[out=120,in=330] node[midway, above, yshift=3pt, sloped] {Coulomb} (C);
\end{tikzpicture}
\caption{The space of supersymmetric deformations of the $\mathrm{SO}(2N+1)$/$\mathrm{USp}(2N)$ $\mathcal{N}=4$ super-Yang-Mills, and the subspace of conformal deformations.}
\label{fig:deformations}
\end{figure}

A second repercussion of our result is that it rules out the conjectured isomorphism between the Coulomb branch cohomology and the monotone cohomology. Indeed, although the Coulomb branch cohomologies of \(\mathrm{SO}(2N{+}1)\) and \(\mathrm{USp}(2N)\) are isomorphic, their monotone cohomologies are not, owing to an additional bosonic monotone class in the \(\mathrm{SO}(7)\) theory with gradings \((1,1,2,2,2,8)\). We have also verified explicitly by (super)abelianizing the 85 monotone classes in this specific sector of the \(\mathrm{SO}(7)\) theory that there is exactly one monotone class not in the Coulomb branch cohomology. It is natural to expect this class to be lifted by higher-loop or non-perturbative effects at the origin of the Coulomb branch. As depicted in Figure~\ref{fig:deformations}, conformal deformations of the complexified gauge coupling $\tau$ and Coulomb branch deformations combine into a larger space of supersymmetric deformations, and the simplest scenario would be that the same monotone class gets lifted in all deformation directions starting from the $\mathrm{SO}(2N+1)$ point.

Third, note that the $(J_1,J_2,q_1,q_2,q_3) = (0,0,3,3,3)$ charge sector belongs to the $\frac18$-BPS chiral ring. It was stated in \cite{Kinney:2005ej} that all cohomology classes in the $\frac18$-BPS chiral ring reside in the Coulomb branch cohomology; our result serves as a counterexample to this statement. The matching of chiral ring operators has historically served as a strong validation of Seiberg dualities \cite{Seiberg:1994pq}, but explicit checks were performed only for operators of small dimensions or low ranks \cite{Romelsberger:2005eg}. 

Finally, one may retrospectively question whether the first fortuitous class in the \(\mathrm{SU}(2)\) theory at level \(L=24\), identified in \cite{Chang:2022mjp}, survives higher-loop and non-perturbative corrections. We can argue that it does. Up to $L=24$, we found that the monotone classes coincide exactly with the Coulomb branch cohomology classes, so there is no additional class with the same charges that could pair with the fortuitous class to get lifted. Whether the monotone classes coincide fully with the Coulomb branch cohomology classes for $\mathrm{SU}(2)$ remains an open question.

\section*{Acknowledgements} 

We thank Jaehyeok Choi, Seok Kim, Zohar Komargodski, and Eunwoo Lee for helpful discussions, and Seok Kim for comments on the draft. CC is partly supported by the National Key R\&D Program of China (NO. 2020YFA0713000). This work was performed in part at the Aspen Center for Physics, which is supported by National Science Foundation grant PHY-2210452.

\bibliography{refs}
\bibliographystyle{JHEP}

\end{document}